\documentstyle[sprocl]{article}

\def\be{\begin{equation}}
\def\ee{\end{equation}}
\def\bea{\begin{eqnarray}}
\def\eea{\end{eqnarray}}

\begin{document}
\title{Phenomenology of Neutrino Mass Matrix
\footnote{Invited Talk at NOON 2000  in December 2000 at Tokyo}
}

\author{N. Haba}

\address{Faculty of Engineering, Mie University, Tsu, Mie, 514-8507,
  Japan \\E-mail: haba@eken.phys.nagoya-u.ac.jp} 

\author{J. Sato}

\address{Research Center for Higher Education, Kyushu University, Fukuoka,
  810-8560 \\E-mail: joe@rc.kyushu-u.ac.jp}

\author{M. Tanimoto}

\address{Department of Physics, Niigata University, Niigata 950-2181,
  Japan\\E-mail: tanimoto@muse.hep.sc.niigata-u.ac.jp}

\author{K. Yoshioka}

\address{Yukawa Institute for Theoretical Physics, Kyoto University, Kyoto
  606-8502, Japan\\E-mail: yoshioka@yukawa.kyoto-u.ac.jp}


\maketitle\abstracts{The search for possible mixing patterns of charged 
leptons and
neutrinos is important to get clues of the origin of nearly maximal
mixings, since there are some preferred bases of the lepton mass
matrices given by underlying theories. We
systematically examine the mixing patterns which could lead to large
lepton mixing angles. We find out 37 mixing patterns are consistent
with experimental data if taking into account phase factors in the
mixing matrices. Only 6 patterns of them can explain the observed data
without any tuning of parameters, while the others need particular
choices for phase values. }
\section{Introduction}

The Super-Kamiokande experiment has confirmed the neutrino oscillation
in atmospheric neutrinos, which favors the
$\nu_\mu\rightarrow\nu_\tau$ process with a large mixing angle 
$\sin^2 2\theta_{\rm atm} \geq 0.88$ and a mass-squared difference
$\Delta m^2_{\rm atm}=(1.6-4)\times 10^{-3}$ eV$^2$~\cite{SKam}. On
the other hand, the recent data of Super-Kamiokande favors the large
mixing angle (LMA) MSW solution~\cite{N2000} for the solar neutrinos
problem, but there are still four solutions allowed;
the small mixing angle (SMA) MSW~\cite{MSW}, the LMA-MSW, the low
$\Delta m^2$ (LOW), and the vacuum oscillation (VO)
solutions~\cite{BKS}. As a result, the neutrino mixing matrix (MNS
 matrix~\cite{MNS}) has two possibilities:
one is the matrix with single maximal mixing, which gives the SMA-MSW
solution for the solar neutrino problem, and the other with bi-maximal
mixing~\cite{bimax}, which corresponds to the LMA-MSW, LOW, and VO
solutions.

Assuming that the neutrino oscillations only account for the solar and
atmospheric neutrino data, one can consider prototypes of the MNS
mixing matrix $U_{\rm MNS}$ with  single maximal mixing, and 
with  bi-maximal mixing. 
Where is the origin of the above nearly maximal mixings? This is one
of the most important problems in the lepton mixing. In almost all
models for the fermion masses and mixing, there are some preferred
bases given by underlying theories of the models. The maximal mixing
angles generally follow from both the charged-lepton and neutrino mass 
matrices. The search for possible mixing patterns of charged leptons
and neutrinos is therefore important for constructing models with
maximal lepton mixings. We systematically investigate
the mixing patterns where at least one of the mixing matrices has
sources of maximal mixings \cite{REF}. Our analysis is not concerned with any
particular structures of lepton mass matrices and hence with the mass
spectrum of neutrinos.
\section{Phenomenology of Mixing Matrices}

When the charged-lepton and neutrino mass matrices are given, the MNS
matrix is defined as
\begin{equation}
  U_{\rm MNS} \,=\, V_E^\dagger\, V_\nu,
\end{equation}
where $V$'s are the mixing matrices which rotate the left-handed
fields so that the mass matrices are diagonalized. The mixing matrices
$V_E$ and $V_\nu$ are generally parameterized as follows:
\begin{eqnarray}
  V_E=P\,U(23)\,P'\,U(13)\,U(12)P'', \qquad
  V_\nu=\overline P\,\overline U(23)\,\overline P'\,\overline U(13)\,
  \overline U(12)\,\overline P''.
\end{eqnarray} 
Here $U(ij)$ are the rotation matrices,
{\small
\begin{equation}
  U(23)= \left( 
    \matrix{ 1 & 0 & 0 \cr
      0 & c_{23} & s_{23} \cr 
      0 & -s_{23} & c_{23} \cr} \right), 
  U(13)= \left( 
    \matrix{ c_{13} & 0 & s_{13} \cr
      0 & 1 & 0 \cr
      -s_{13} & 0 & c_{13} \cr} \right),
  U(12)= \left( 
    \matrix{ c_{12} & s_{12} & 0 \cr
      -s_{12} & c_{12} & 0 \cr
      0 & 0 & 1 \cr} \right)
\end{equation}}
in which $s_{ij}=\sin\theta_{ij}$ and $c_{ij}=\cos\theta_{ij}$, 
and $P$'s are the phase matrices; $P={\rm diag} (1, e^{ia}, e^{ib})$, 
$P'={\rm diag} (1, 1, e^{i\delta})$, and 
$P''={\rm diag} (e^{ip}, e^{iq}, e^{ir})$. The matrices 
$\overline U(ij)$, $\overline P$, $\overline P'$, and $\overline P''$
in the neutrino side take the same forms as above. 
Since there are six mixing angles in $V_E$ and $V_\nu$, it
is meaningful to raise a query which angles are responsible for the
observed maximal mixings in $U_{\rm MNS}$. In order to answer this, we
analyze the mixing patterns in a model-independent way.

Now, the MNS mixing matrix is written as 
\begin{eqnarray}
  U_{\rm MNS} = \{U(23)\,P'\,U(13)\,U(12)\}^\dagger\, Q\,
  \overline U(23)\,\overline P'\,\overline U(13)\,\overline U(12) 
  \equiv U_E^\dagger\,Q\,U_\nu,
\end{eqnarray} 
in which
  $Q = P^*\overline P \equiv diag(1, e^{i\alpha}, e^{i\beta})$.
As will be seen below, in our analysis, the phase factors in the
matrix $Q$ sometimes play important roles to have phenomenologically
viable mixing angles. The mixing matrix $U(ij)$ and $\overline U(ij)$
are determined if the mass matrices of charged leptons and neutrinos
are given. In the first approximation, we assume that these mixing
angles are zeros or maximal ones, and then examine possible
combinations of $U_E$ and $U_\nu$ combined with indications of 
Super-Kamiokande and long baseline neutrino experiments.

Let us consider 9 types of mixing matrices for $U_E$ and $U_\nu$. The
first three types of matrices are given by taking one of mixing angles
being maximal and the others being zero:
\begin{eqnarray}
  A = \left( 
    \matrix{ 1 & 0 & 0 \cr
      0 & \frac{1}{\sqrt{2}} & \frac{1}{\sqrt{2}} \cr
      0 & -\frac{1}{\sqrt{2}} & \frac{1}{\sqrt{2}} \cr} \right), \
   S = \left( 
    \matrix{ \frac{1}{\sqrt{2}} & \frac{1}{\sqrt{2}} & 0 \cr
      -\frac{1}{\sqrt{2}} & \frac{1}{\sqrt{2}} & 0 \cr
      0 & 0 & 1 \cr} \right), \
    L = \left(
    \matrix{ \frac{1}{\sqrt{2}} & 0 & \frac{1}{\sqrt{2}} \cr
      0 & 1 & 0 \cr
      -\frac{1}{\sqrt{2}} & 0 & \frac{1}{\sqrt{2}} \cr} \right)
 \end{eqnarray} 
where we use the notation $A$, $S$, and $L$ for three type mixing
matrices of $U_E$ and $U_\nu$. The second three types of matrices are
given by taking one of mixing angles being zero and the others being
maximal:
\begin{eqnarray}
  B = \left(
    \matrix{ \frac{1}{\sqrt{2}} & \frac{1}{\sqrt{2}} & 0 \cr
      -\frac{1}{2} & \frac{1}{2} & \frac{1}{\sqrt{2}} \cr
      \frac{1}{2} & -\frac{1}{2} & \frac{1}{\sqrt{2}} \cr} \right), 
  H = \left(
    \matrix{ \frac{1}{2} & \frac{1}{2} & \frac{1}{\sqrt{2}} \cr
      -\frac{1}{\sqrt{2}} & \frac{1}{\sqrt{2}} & 0 \cr
      -\frac{1}{2} & -\frac{1}{2} & \frac{1}{\sqrt{2}} \cr} \right),
  N = \left(
    \matrix{ \frac{1}{\sqrt{2}} & 0 & \frac{1}{\sqrt{2}} \cr
      -\frac{1}{2} & \frac{1}{\sqrt{2}} & \frac{1}{2} \cr
      -\frac{1}{2} & -\frac{1}{\sqrt{2}} & \frac{1}{2} \cr} \right) . 
 \end{eqnarray} 
The threefold maximal mixing~\cite{threefold} and the unit matrix are
also added in our analyses:
\begin{eqnarray}
  T = \left(
    \matrix{ \frac{1}{\sqrt{3}} & \frac{1}{\sqrt{3}} &
      \frac{1}{\sqrt{3}} e^{-i\delta} \cr
      -\frac{1}{2}-\frac{1}{2\sqrt{3}}e^{i\delta} & \frac{1}{2}
      -\frac{1}{2\sqrt{3}}e^{i\delta} & \frac{1}{\sqrt{3}} \cr
      \frac{1}{2}-\frac{1}{2\sqrt{3}}e^{i\delta} & -\frac{1}{2}
      -\frac{1}{2\sqrt{3}}e^{i\delta} & \frac{1}{\sqrt{3}} \cr}
    \right),\quad
   I &=& \left( 
    \matrix{ 1 & 0 & 0 \cr
      0 & 1 & 0 \cr
      0 & 0 & 1 \cr} \right).
 \end{eqnarray}
In addition to these, one specific mixing, which is the so-called
democratic type mixing~\cite{Democratic}, is examined because this
mixing is different from the above ones and may be derived from
well-motivated underlying theories:
\begin{eqnarray}
  D &=& \left(
    \matrix{ \frac{1}{\sqrt{2}} & \frac{1}{\sqrt{6}} &
      \frac{1}{\sqrt{3}} \cr
      -\frac{1}{\sqrt{2}} & \frac{1}{\sqrt{6}} & \frac{1}{\sqrt{3}}\cr 
      0 & -\frac{2}{\sqrt{6}} & \frac{1}{\sqrt{3}} \cr} \right). 
  \quad\qquad
  \matrix{ s_{12}=1/2 \hfill\cr
    s_{13}=1/\sqrt{3} \hfill\cr
    s_{23}=1/\sqrt{2} \hfill\cr}
\end{eqnarray}

By using the above types of mixing matrices, we have 81 ($=9\times 9$) 
combinations of matrices for the MNS matrix $U_{\rm MNS}$, in which
the phases $\alpha$, $\beta$, $\delta_E$, and $\delta_\nu$ are free
parameters. Note that if at least one of the matrix elements is zero
in $U_E$ ($U_\nu$), we can take $P'$ ($\overline P'$) as a unit matrix
without loss of generality. The phase $\delta_E$ ($\delta_\nu$) can be
absorbed into the matrices $P''$ ($\overline P''$) and/or $Q$. 

We examine the MNS matrices according to phenomenological constraints
coming from the atmospheric neutrino experiments. The Chooz
experiment~\cite{CHOOZ} also provides a useful guide for the
classification of mixing matrices, in particular, for the 
$(U_{\rm MNS})_{e3}$ element. The solar neutrino problem can be solved
with both large and small mixing angle solutions, which are now
predictions of our systematic search of taking 81 combinations of
$U_E$ and $U_\nu$. We here take a convention where the mixing between
the labels 2 and 3 is relevant to atmospheric neutrinos and the mixing
between the labels 1 and 2 to the solar neutrino problem. We find that
the 81 mixing patterns are classified into the following five categories:
\begin{itemize}
\setlength{\itemsep}{0mm}
\item class 1: small mixing for atmospheric neutrinos 
\item class 2: large value of $(U_{\rm MNS})_{e3}$ 
\item class 3: small mixing for atmospheric neutrinos when 
  $(U_{\rm MNS})_{e3}\ll 1$ by tuning phase values
\item class 4: consistent with experiments by tuning phase values
\item class 5: consistent with experiments independently of phase
  values
\end{itemize}
The classes 4 and 5 are consistent with the experimental data.
 We have also
checked the ``stability'' of our classification numerically by taking
the fluctuations of all mixing angles in the region of
$\theta_{ij}=\theta_{ij}\pm 5^\circ$ both in the charged-lepton and
neutrino sectors.
 Due to a constraint from the Chooz experiment,
one may usually assume that a bi-maximal mixing matrix takes the form
of type $B$. It is, however, found here that the matrix $N$, which has
a large 1-3 mixing, gives exactly the same results as the matrix $B$
does. (The predictions for 1-2 mixing angles are also the same.) This
would give a new possibility of model-building for the fermion masses
and mixings.

In the category of class 5, there are the following 6 mixing patterns: 
\begin{equation}
  (U_E,\,U_{\nu}) \;=\;
  (A,\,S),\;\; (A,\,I),\;\; (I,\,A),\;\; (I,\,B),\;\; (D,\,S),\;\;
  (D,\,I).
  \label{class5}
\end{equation}
 An interesting fact we find in (\ref{class5}) is that
without tuning of phase parameters (i.e., in class 5), a large 1-2
mixing relevant to the solar neutrino problem must come from the
neutrino side (except for the cases of democratic-type mixing). This
may be a natural result in view of the charged-lepton mass matrix and
commonly discussed in the literature. That is, in the charged-lepton
sector, the mass hierarchy between the first and second generations
may be too large for the large angle solar solutions.
  It is, however, noted that the same
result can be obtained only from a viewpoint of mixing matrices.

Class 4 contains the other 31 patterns of mixing matrices. These
patterns require suitable choices of phase values to be consistent
with the experimental data. The result is summarized in Table 1, which
shows the values of mixing angle for atmospheric neutrinos 
($\sin^2 2\theta_{\rm atm}$) and for solar neutrinos 
($\sin^2 2\theta_\odot$) in cases that the values of 
$(U_{\rm MNS})_{e3}$ are fixed to be minimum. For each combination, we
also present the relevant phases which are tuned to obtain the minimum
value of $(U_{\rm MNS})_{e3}$. In some cases, the mixing angles of
$\sin^2 2\theta_{\rm atm}$ and $\sin^2 2\theta_\odot$ have some
uncertainties in their predictions. It is because there are remaining
phase degrees of freedom even with fixed values of 
$(U_{\rm MNS})_{e3}$. The mixing patterns in class 4 have different
numbers of phase tuning to obtain experimentally suitable MNS
matrices. For example, the types $(U_E,\,U_\nu)=(A,\,A)$ and 
$(A,\,B)$, which are often seen in the literature, requires only one
tuning of phase values to fix all the mixing angles in $U_{\rm MNS}$
(see also the next section).
These combinations have not been discussed so
far in the literature and would provide new possibilities for
constructing models where fermion masses and mixing angles are
properly reproduced.
\section{Texture of Lepton Mass Matrices}

Let us begin with discussing the patterns in class 5. As noted in the
previous section, these mixing patterns are well known in the
literature, in other words, there are a lot of models of mass matrices
which lead to these mixing patterns. We summarize the 6 patterns
briefly assuming that  neutrinos are Majorana.

The first pattern is the case $(U_E,\,U_\nu)=(A,\,S)$, which
predicts bi-maximal mixing for the MNS matrix.
 Such a type of texture is at first derived in SO(10) grand
unified models~\cite{P1}.
 The next one is the case  $(U_E,\,U_\nu)=(A,\,I)$,
 which predicts single maximal mixing for the MNS matrix.
These mass matrices are indeed obtained in $E_7$, $E_6$, and $SO(10)$
grand unified theories~\cite{P2}. The third one is
$(U_E,\,U_\nu)=(I,\,A)$, which gives single maximal mixing for the MNS
matrix.
This texture can be given by, for example, $R$-parity violating
models~\cite{P3}. The fourth one is $(U_E,\,U_\nu)=(I,\,B)$, which
gives bi-maximal mixing for the MNS matrix.
This texture follows from radiative generation mechanisms for neutrino 
masses~\cite{P4}.
The fifth and sixth patterns are specific ones because they depend on
the democratic lepton mass matrix~\cite{Democratic}.
All the above mixing patterns are allowed by the experimental
data without any tuning of phases, $\alpha$, $\beta$, $\delta_{E,\,\nu}$.

Next let us discuss the mixing patterns in class 4, where the presence 
of phase factors is essential in the MNS matrix to have right values
of mixing angles. There are 31 patterns classified into this category,
but only a few mass matrix models with these mixing patterns have been
proposed. These patterns thus provide potentially useful
textures of lepton mass matrices. 

At first, we discuss a well-known example $(U_E,\,U_\nu)=(A,\,A)$
which is derived from the mass matrices,
\begin{eqnarray}
  M_E \,\propto\, \left( 
    \matrix{ & & \cr
      & \lambda^2 & 1 \cr
      & \lambda^2 & 1 \cr} \right), \qquad  
  M_\nu \,\propto\, \left(
    \matrix{ & & \cr
     & 1 & 1 \cr
     & 1 & 1 \cr} \right),
\end{eqnarray} 
obtained in the models with $U(1)$ flavor symmetries~\cite{U1,vissani}. The
mixing angles at leading order become
\begin{eqnarray}
  \theta_{\rm atm} \,=\, \frac{\beta-\alpha}{2},\qquad
  \theta_\odot \,=\, (U_{\rm MNS})_{e3} \,=\, 0,
  \label{E41}
\end{eqnarray}
that gives a SMA solution for the solar neutrino problem. The
experimental constraint, $(U_{\rm MNS})_{e3}\ll 1$, is satisfied, but
for atmospheric neutrinos, tuning of phase values for $\beta-\alpha$
must be involved. In the presence of the phase matrix $Q$, a
cancellation of two large mixing angles from $U_E$ and $U_\nu$ can be
avoided.

Another pattern for which concrete models have been constructed is the
case of $(U_E,\,U_\nu)=(A,\,B)$, which can be derived from the mass
matrices,
\begin{eqnarray}
  M_E \,\propto\, \left( 
    \matrix{ & & \cr
      & \lambda^2 & 1 \cr
      & \lambda^2 & 1 \cr} \right), \qquad  
  M_\nu \,\propto\, \left( 
    \matrix{ & 1 & 1 \cr
      1 & & \epsilon \cr
      1 & \epsilon & \cr} \right).
  \label{P42}
\end{eqnarray}
These textures have been discussed in Ref.~\cite{Shafi}. It is pointed
out in Ref.~\cite{vissani} that this combination of the mixing
matrices is also derived from the mass matrices in Eq.~(\ref{E41}).  The
mixing angle $\theta_{\rm atm}$ is the same as Eq.~(\ref{E41}), and a
phase value, $\beta-\alpha\simeq \pi/2$ must be chosen to get
maximal mixing of atmospheric neutrinos.

We here comment on the models~\cite{BB} which introduce the following
type of mass matrices: 
\begin{eqnarray}
  M_E \,\propto\, \left( 
    \matrix{\lambda^3 & \lambda^2 & 1 \cr
    \lambda^3  & \lambda^2 & 1 \cr
    \lambda^3  & \lambda^2 & 1 \cr} \right), \qquad  
  M_\nu \,\propto\, \left(
    \matrix{1 & 1 & 1 \cr
    1 & 1 & 1 \cr
    1 & 1 & 1 \cr} \right). 
\end{eqnarray} 
This corresponds to $(U_E,\,U_\nu)=(T,\,T)$, or to a special case 
$(U_E,\,U_\nu)=(D,\,D)$, where suitable MNS matrices can also be
obtained by phase tuning.

As noted in the previous section, in class 4, there are several new
mixing patterns which have not yet been discussed. 
 Let us show an example of
the case $(U_E,\,U_\nu)=(S,\,N)$. This mixing pattern could be derived
from 
\begin{eqnarray}
  M _E \,\propto\, \left( 
    \matrix{ & \lambda^4 & \lambda^2 \cr
      & \lambda^4 & \lambda^2 \cr
      & \lambda^2 & 1 \cr} \right), \qquad  
  M_\nu \,\propto\, \left( 
    \matrix{ 2 & \sqrt{2} & \sqrt{2} \cr
      \sqrt{2} & 1+\epsilon & 1-\epsilon \cr
      \sqrt{2} & 1-\epsilon & 1+\epsilon \cr} \right) \ .
  \label{P43}
\end{eqnarray} 
In this case, we have 
\begin{eqnarray}
   \sin^2 2\theta_{\rm atm} \,=\, \sin^2 2\theta_\odot 
  \,=\, \left|\frac{1}{2}+\frac{1}{2\sqrt{2}}e^{i\alpha}\right|^2, \ \ 
   (U_{\rm MNS})_{e3} \,=\, 
   \left|\frac{1}{2}-\frac{1}{2\sqrt{2}}e^{i\alpha}\right|.
  \label{P43m}
\end{eqnarray}
Here we would like to emphasis that a single phase tuning of $\alpha$
can ensure all the mixing angles to be consistent with
experiments. Since the
$(U_{\rm MNS})_{e3}$ mixing in (\ref{P43m}) is close to the Chooz
bound, this pattern will be tested in the near future. Including the
above example, we find several possible mixing patterns which no one
has discussed so far (see Table 1). Model-construction utilizing such
types of textures may be worth being performed.
\section{Summary and Discussion}

We have found that there are many allowed mixing patterns of charged
leptons and neutrinos for the MNS matrix with bi-maximal or single
maximal mixing. Among them, only 6 mixing patterns are allowed without
any tuning of phase values. Interestingly, these patterns are indeed
derived from the concrete models which have been proposed to account
for the fermion mass hierarchy problem. The other patterns can give
solutions of the observed neutrino anomalies depending on the choices
of phase values. In this class, physically significant mixing patterns
might be the ones which need a fewer numbers of phase tuning to have
definite predictions consistent with experiments. We have found that
9 combinations satisfy this criterion, a single phase tuning
required. They have not been studied enough in mass matrix models and
would give new possibilities of model-construction. The phases to be
tuned are not completely unphysical unlike the quark sector, but some
of them could be connected to Majorana phases and $CP$ violation in
the lepton sector. Combined with these effects,  the
measurements of mixing angles $\sin^2 2\theta_\odot$ and 
$(U_{\rm MNS})_{e3}$ will be important to select possible mixing
patterns.

\subsection*{Acknowledgments}

 We would like to thank T. Yanagida for useful discussions and comments.
 We also  thank the organizers and participants of Summer 
Institute 2000 held at Yamanashi, Japan, where a portion of this work 
was carried out. This work is supported by the Grant-in-Aid 
for Science Research, Ministry of Education, Science and Culture, 
Japan (No.~10640274, No.~12047220, No.~12740146, No.~12014208).

\begin{table}
\begin{center}
\def\arraystretch{2}
\tabcolsep=8pt
\begin{tabular}{|r|r|r|r|r|} \hline
  {$U_E$|$U_\nu$} & $\sin^2 2\theta_{\rm atm}$ & 
  ~~~~$\sin^2 2\theta_\odot$~~  & $(U_{\rm MNS})_{e3}$ & 
  {\small (\# of) phases} \\ \hline\hline
  {$A$|$A$} & $0-1$ & 0 & 0 & (0) \\
  {$A$|$B$} & $0-1$ & 1 & 0 & (0) \\
  {$S$|$D$} & 8/9 & 0 & 0 & $\alpha$ \ (1) \\
  {$S$|$T$} & 8/9 & $1/4-1$ & 0 & $\alpha+\delta_\nu$ \ (1) \\
  {$S$|$N$} & 0.73 & 0.73 & 0.15 & $\alpha$ \ (1) \\
  {$L$|$T$} & 8/9 & $1/4-1$ & 0 & $\beta+\delta_\nu$ \ (1) \\
  {$L$|$N$} & 0.73 & 0.73 & 0.15 & $\beta$ \ (1) \\
  {$L$|$D$} & 8/9 & 3/4 & 0 & $\beta$ \ (1) \\
  {$B$|$T$} & 8/9 & $1/4-1$ & 0 & $\alpha$, \ $\beta$ \ (2) \\
  {$B$|$H$} & 0.73 & $0.23-0.96$ & 0.15 & $\beta$ \ (1) \\
  {$B$|$L$} & 0.73 & 0.73 & 0.15 & $\beta$ \ (1) \\
  {$B$|$D$} & 8/9 & 15/16 & 0 & $\alpha$, \ $\beta$ \ (2) \\
  {$H$|$T$} & 8/9 & $1/16-1$ & 0 & $\alpha$, \ $\beta$ \ (2) \\
  {$H$|$B$} & 0.73 & $0.23-0.96$ & 0.15 & $\alpha-\beta$ \ (1) \\
  {$H$|$N$} & 1 & 1 & 0 & $\alpha$, \ $\beta$ \ (2) \\
  {$H$|$A$} & 0.73 & 0.73 & 0.15 & $\alpha-\beta$ \ (1) \\
  {$H$|$D$} & 8/9 & 15/16 & 0 & $\alpha-\beta$ \ (1) \\ \hline
\end{tabular}
\caption{The mixing patterns in class 4. The values of mixing angles 
are shown in case of $(U_{\rm MNS})_{e3}$ being minimal. The last 
column denotes the (number of) relevant phases which are needed for
tuning $(U_{\rm MNS})_{e3}$. The uncertainties in 
$\sin^2 2\theta_{\rm atm}$ and $\sin^2 2\theta_\odot$ are fixed by
additional phase tunings.}
\end{center}
\end{table}

\begin{table}
\begin{center}
\def\arraystretch{2}
\tabcolsep=8pt
\begin{tabular}{|r|r|r|r|r|} \hline
  {$U_E$|$U_\nu$} & $\sin^2 2\theta_{\rm atm}$ & 
  ~~~~$\sin^2 2\theta_\odot$~~  & $(U_{\rm MNS})_{e3}$ & 
  {\small (\# of) phases} \\ \hline\hline
  {$N$|$T$} & 8/9 & $1/4-1$ & 0 & $\alpha$, \ $\beta$ \ (2) \\
  {$N$|$H$} & 0.73 & $0.23-0.96$ & 0.15 & $\beta$ \ (1) \\
  {$N$|$L$} & 0.73 & 0.73 & 0.15 & $\beta$ \ (1) \\
  {$N$|$D$} & 8/9 & 15/16 & 0 & $\alpha$, \ $\beta$ \ (2) \\
  {$T$|$T$} & $0-1$ & $0-1$ & 0 & $\alpha+\delta_\nu$, \ 
  $\beta+\delta_\nu$ \ (2) \\
  {$T$|$B$} & 1 & $1/9-1$ & 0 & $\delta_E$, \ $\alpha-\beta$ \ (2) \\
  {$T$|$H$} & 1 & $1/9-1$ & 0 & $\delta_E$, \ $\beta$ \ (2) \\
  {$T$|$N$} & $8/9-1$ & 8/9 & 0 & $\alpha$, \ $\beta$ \ (2) \\
  {$T$|$A$} & 1 & 8/9 & 0 & $\delta_E$, \ $\alpha-\beta$ \ (2) \\
  {$T$|$L$} & 1 & 8/9 & 0 & $\delta_E$, \ $\beta$ \ (2) \\
  {$T$|$D$} & $0-1$ & $0-1$ & 0 & $\alpha$, \ $\beta$ \ (2) \\
  {$D$|$T$} & $0-1$ & $1/4-1$ & 0 & $\alpha+\delta_\nu$ \ (1) \\
  {$D$|$N$} & $1/36-0.96$ & 0.73 & 0.15 & $\alpha$ \ (1) \\
  {$D$|$D$} & $0-1$ & 0 & 0 & $\alpha$ \ (1) \\ \hline
\end{tabular}
\vskip 3mm

Table 1 (continued.)
\end{center}
\end{table}

\end{document}